# Topology-optimized ultra-compact all-optical logic devices on silicon photonic platforms

Lu He[1],† Furong Zhang[1],† Huizhen Zhang[1], Ling-Jun Kong[1], Weixuan Zhang[1], Xingsheng Xu[2], and Xiangdong Zhang[1]*

[1]*Key Laboratory of advanced optoelectronic quantum architecture and measurements of Ministry of Education, Beijing Key Laboratory of Nanophotonics; Ultrafine Optoelectronic Systems, School of Physics, Beijing Institute of Technology, 100081 Beijing, China.*

[2]*State Key Laboratory of Integrated Optoelectronics, Institute of Semiconductors, Chinese Academy of Sciences, Beijing 100083, People's Republic of China.*

†*These authors contributed equally to this work.* * Author to whom any correspondence should be addressed: *zhangxd@bit.edu.cn*.

## Abstract

**The realization of all-optical integration and optical computing has always been our goal. One of the most significant challenges is to make integrated all-optical logic devices as small as possible. Here, we report the implementation of ultra-compact all-optical logic devices and integrated chips on silicon photonic platforms by topology optimization. The footprint for the fabricated all-optical logic gates with XOR and OR functions is only 1.3×1.3 μm² (~0.84λ×0.84λ), that are the smallest all-optical dielectric logic devices ever verified in experiments in the optical communication range. The ultra-low loss of the optical signal is also demonstrated experimentally (-0.96dB). Furthermore, an integrated chip containing seven major logic gates (AND, OR, NOT, NAND, NOR, XOR, and XNOR) and a half adder is fabricated, where the associated footprint is only 1.3×4.5 μm². Our work opens up a new path towards practical all-optical integration and optical computing.**

# 1. Introduction

Logic gates are essential components of the Central Processing Unit (CPU) in a computer. Similarly, all-optical logic gates play key roles in all-optical networks, ultrafast optical computing, and signal processing[1-3]. It is expected that integrated chips could contain as many all-optical logic gates as possible to fulfill various functionalities in the information processing. This requires the fabricated logic gate to be as small as possible. In order to fabricate these devices, various technologies and materials have been introduced for decades[4-15].

Traditionally, all-optical logic gates are constructed by using nonlinear optical fibers[16-17] and semiconductor optical amplifiers[18-20]. While, the large size of these conventional apparatus makes the proposed all-optical logic gates cannot be integrated. The footprints of plasmonic logic gates are much smaller than the wavelength scale. However, they have a non-negligible weakness, the large optical loss, which makes them very difficult to be used in the integrated chip[21-23]. In contrast, photonic-integrated circuits fabricated on the silicon-on-insulator (SOI) platform have attracted substantial attentions due to the low power consumption and high transmission efficiency. Traditional optical logic devices based on the SOI still possess a large size, that cannot satisfy the requirement of modern on-chip integrated optical systems[24-30]. Recently, the ultra-short and low-loss $\Psi$ gate of the linear optical logic has been prepared on silicon (Si) photonic platforms with several micrometers[31]. However, such a $\Psi$ gate still has a footprint of about double wavelengths. The question is whether we can create all-optical dielectric logic devices with extremely smaller sizes, even smaller than the wavelength scale, with multiple functionalities.

On the other hand, recent investigations have shown that the method of topology optimization (TO) could display various advantages in the design of compact optoelectronic devices[31-40], and many basic elements have been designed, including beam splitters[37], polarization beam splitters[38, 39], photonic cavities[40], and so on. These proposed devices have better performances and more compact structures. It is meaningful to ask whether much more exotic effects could appear when the TO method is applied to the design of all-optical logic devices on a chip.

In this work, we design all-optical logic devices on Si photonic platforms using the TO method. In particular, the all-optical dielectric logic gates with the sub-wavelength size are realized. For example, the footprint of XOR and OR logic gates is only 1.3×1.3 μm$^2$ (~0.84λ×0.84λ) around the optical communication wavelength. The size of a chip including 7 major logic gates (AND, OR, NOT, NAND, NOR, XOR, and XNOR) and a half adder is only 1.3×4.5 μm$^2$. Furthermore, we fabricate these super-compact all-optical logic devices and chips. The experimental results show that these devices have the ultra-low loss and high-efficient performances. It is expected that our optimized all-optical logic devices have potential applications in the field of all-optical networks, ultrafast optical computing, and signal processing.

## 2. The theoretical model of the linear all-optical logic gates.

The all-optical logic gates we designed are based on the linear interference approach[28-31]. In our model, several 50:50 beam splitters (BSs) are used. As shown in Fig. 1a, the fundamental logic unit (XOR/OR gate) is made by only one BS. When two signals with the phase difference $\Delta\varphi_{AB}=\pi/2$ are injected from two input ports at the same time, the signal can only go out from the OR port owing to the complete interference cancellation at the XOR port. In addition, if only one signal is injected into the system from any input port (A or B), the output fields could leave from the system at two output ports (XOR and OR ports) with equal amplitudes at the same time. That is to say, the XOR and OR gates are successfully realized.

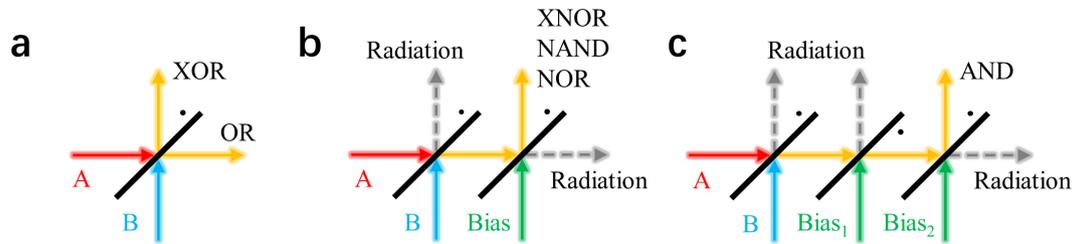

**Fig. 1. The principle of linear optical logic gates by 50:50 BSs. a,** The XOR and OR gates by one 50:50 BS. **b,** The XNOR, NAND, and NOR gates by two 50:50 BSs. **c,** The AND gate by three 50:50 BSs. The black line represents the 50:50BS. The black dot marks the side of destructive interference. Red and blue arrows represent Inputs A and B. Yellow arrows represent the output signals. Gray dotted-line arrows represent the radiation signals.

And then in Fig. 1b, we show the cascaded logic gates, including XNOR, NAND and NOR gates, by two BSs. The output signal of the OR gate (from the left BS) and the bias light (always turn-on) are injected into the right BS. The phase difference between them is π, so destructive interference happens on the output port. Particularly, the amplitudes of bias lights for XNOR, NAND, and NOR gates are bias=$\sqrt{2}/2$, $\sqrt{2}$, and $0.75\sqrt{2}$. At the same time, the different logic gates can be switched by changing the amplitudes of the bias light. To further realize the AND gate, we add a NOT gate after the NAND gate, as shown in Fig. 1c. The NAND output signal and the second bias light (bias$_2$=0.75) are injected in the third BS. Similarly, the phase difference between them is π, so destructive interference happens on the output port. That is, the cascaded logic gates can be realized by using two or three BSs.

The power ratio between logic output 1 and 0 is an important parameter of optical logic gates. In our proposed logic gates, the power ratios of XOR, OR, XNOR, and NAND gates are infinity. As for NOR and AND gates, the power ratios are 9:1. Additionally, it should be pointed that our designed logic gates have some radiation signals at the other ports of the BSs, which is represented by the gray dotted-line arrows in Figs. 1a-1c.

## 3. The OR, XOR, and NOT gates by topology optimization

We use a fully etched 220nm-thick Si layer coated on a SiO$_2$ substrate to design all-optical linear logic gates by TO. It is worthy to note that a 2D Si layer with the effective permittivity could be used to optimize such a 3D structure with high precision. In S1 of Supporting Materials, the method to calculate the effective permittivity of the Si layer with different thicknesses is discussed. At first, we focus on the optimization of OR and XOR gates in a single structure, where a pair of input ports and two output ports (named OR and XOR ports) exist, as shown in Fig. 2a. That is to say, a 50:50 beam splitter (BS) with 2×2 ports can be used to realize the XOR and OR gates. So, the logic gate design changes to the design problem of a simple 50:50 BS.

To fulfill these operations, the system must possess a mirror symmetry with respect the dashed line (called mirror symmetry line), as marked in Fig. 2a. In this case, only a half region is needed to be optimized where the other mirror symmetric part should have the same dielectric distribution.

To achieve the above-mentioned logic gates by TO, the associated objective function with the excitation coming from a single input port (input port A) is defined as:

$$\max_{\varepsilon^{\bullet}_{air} \leq \varepsilon \leq \varepsilon^{\bullet}_{si}} \Phi_{Total}(\varepsilon) = \sum_{\lambda}(I_{XOR} + I_{OR}) \ , \tag{1}$$

where $\varepsilon(\mathbf{r}) \in [\varepsilon^{\bullet}_{air}, \varepsilon^{\bullet}_{si}]$ is the design field (the dielectric distribution of the structure) with the effective permittivities of air and Si layer being $\varepsilon^{\bullet}_{air}=1$ and $\varepsilon^{\bullet}_{si}=8.04$, and $I_{XOR}$ ($I_{OR}$) represents the output intensity from the XOR (OR) port. The λ is the wavelength of the incident field, which is summed in the objective function with three different values (λ=1520nm, 1550nm, 1580nm) to extend the range of operation frequencies. To realize the complete interference cancellation under the double ports' excitations with $\Delta\varphi_{AB}=\pi/2$, the output fields at XOR and OR ports should satisfy the relationship of $E_{out}(XOR)=E_{out}(OR)e^{i\pi/2}$. It is worthy to note that the phase difference between two output ports is always π/2 that is protected by the time-reversal symmetry. Hence, the extra condition $\gamma_1 \leq I_{XOR}/I_{OR} \leq \gamma_2$ should be added in the TO to limit amplitudes of output fields from two ports becoming nearly identical, so that the complete interference cancellation in the optimized 3D structure could be realized. Here, we set $\gamma_1=0.6$ and $\gamma_2=0.7$ in the TO process. In S2 of Supplementary Materials, we give detailed discussions about the influence of $\gamma_1$ and $\gamma_2$ on the performance of the 3D structure by TO. Moreover, to ensure ideal connectivity of the optimized structure, that is the ratio between the area of Si and air should be larger than a threshold. For this purpose, another restriction in the TO is expressed as: $\iint_{Design\ region}(\varepsilon-\varepsilon_{air}^{\bullet})/(\varepsilon_{si}^{\bullet}-\varepsilon_{air}^{\bullet})dS \geq \gamma_3$, where we set $\gamma_3=0.6$ to make the area of Si larger than air. Except for these limiting conditions, the linear material interpolation, projection, and filtering procedure are used in the process of TO solved by COMSOL[36].

As illustrated in Fig. 2b, we display the dielectric distribution of the optimized structure with the iteration step being 200, 600, 800, and 1000 times. It starts from the homogeneous initial value, in which the dielectric constant is $\varepsilon=(\varepsilon_{air}^{\bullet}+\varepsilon_{si}^{\bullet})/2$. With the increasing of iteration steps, the objective function is gradually enlarged and the final configuration with the optimal performance appears after 1000 iterations. We note that there are some narrow connections in the optimized structure, which are hard to be fabricated in experiments. Hence, a few artificial modifications, that do not change the performance of the device, are applied to obtain the executable structure (See detailed discussions in S3 of Supplementary Materials). The optimized 3D logic device is presented in Fig. 2c. The footprint for such a device is only 1.3×1.3 μm$^2$. The detailed discussions on the size of the device are described in S4 of Supporting Materials. In previous works[37, 41-43], lots of compact beam splitters (BSs) are fabricated by the inverse design method. Some of them (with 2 input ports and 2 output ports) can also be used to realize logic gates. However, in the property of footprints and insertion losses, they are not good enough to be used for low-loss and compact logic gates. In Appendix A, we list a comparison table about various BSs[37, 41-43], directional coupler[44, 45] (DC), and multi-mode interference (MMI) couplers[46] based on inverse design and traditional methods.

And then, Fig. 2d shows the simulation results of the steady-state intensity distribution $|\text{Re}(H_z)|^2$ in the optimized structure under a single port and double ports excitations. When one of input ports (A or B) is excited, the logic signals become 1 at XOR and OR ports. And, when incident fields are injected into both input ports (A and B) simultaneously, the output field at the XOR (OR) port becomes 0 (1), owing to the destructive (constructive) interference. Based on these results, we can see that the designed structure by TO possesses a nice performance to implement XOR and OR logic gates. In addition, the NOT gate can also be realized at the XOR port by regarding the input port B as a bias input (always turn-on).

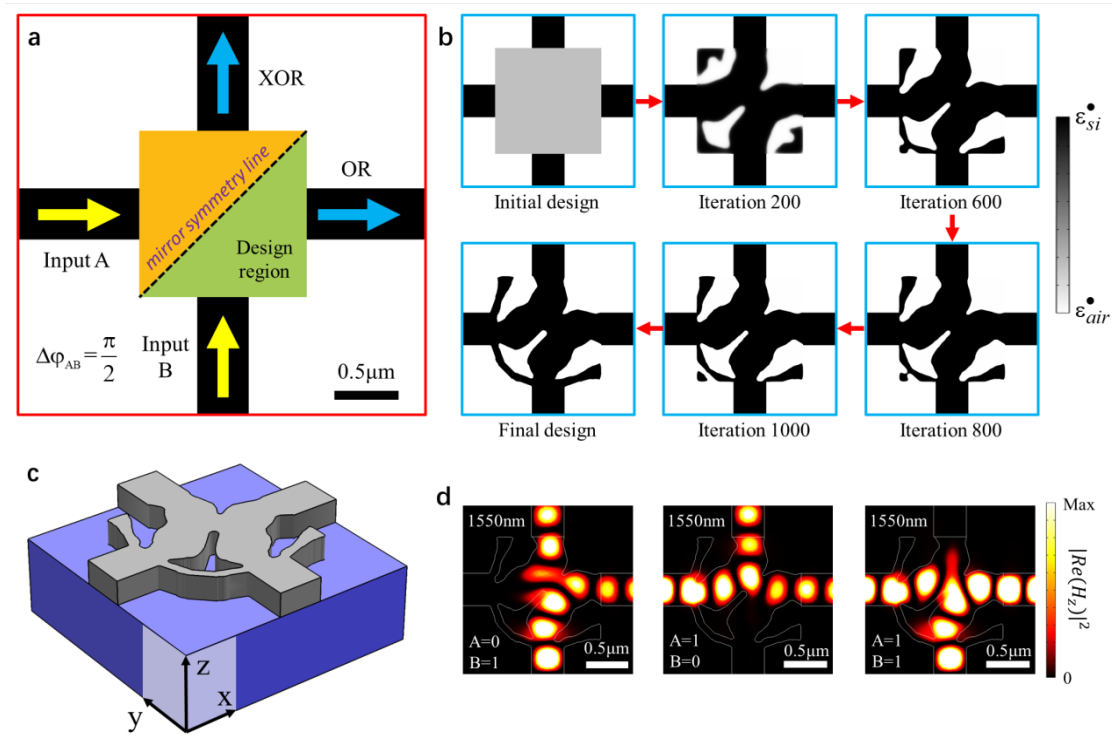

**Fig. 2. The design process of XOR and OR gates by the TO and its numerical simulation results in 3D. a**, The schematic diagram of the optimization model. The designable region (green triangle) and the mirror symmetry line (dashed line). Two input signals (from Input A and B ports) are guided in left and bottom waveguides, and two output signals (from XOR and OR ports) are guided in top and right waveguides. **b**, The process of topology optimization. The initial design is an intermediate material between Si and air. With the optimization going on, the materials in the design region tend to binarization. The final design is determined by 1000 iterations and some artificial modifications. **c**, The 3D view of the logic device. The gray structure represents Si and the blue represents the $SiO_2$. **d**, The simulated steady-state intensity distribution of the XOR and OR gates at λ=1550nm. The pictures correspond to the three logic cases, inputs 01, 10, and 11, respectively.

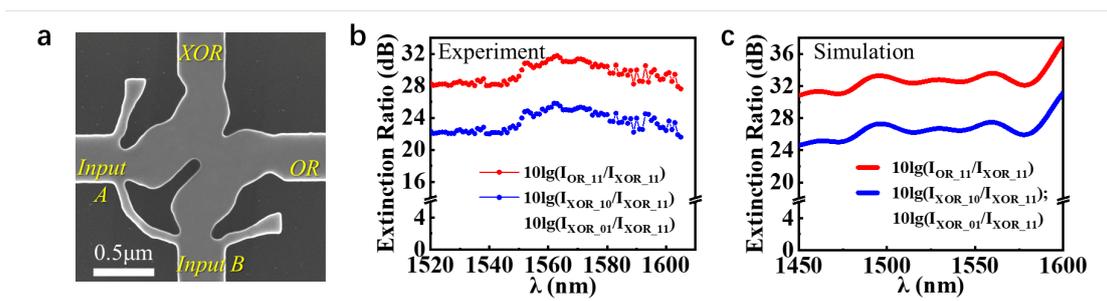

**Fig. 3. The SEM image and ERs of XOR and OR logic gates. a**, The SEM image of the logic device for the top view. **b**, The ERs of the measured logic gate, between logic 1 (OR_11, XOR_01, XOR_10) and logic 0 (XOR_11). **c**, The ERs in simulation corresponding to (**b**).

The designed logic devices are fabricated using electron-beam lithography followed by dry etching (see Appendix B for details). The scanning electron microscopy (SEM) image of the top view is displayed in Fig. 3a. Comparing the fabricated structure with the theoretical design, we find that they show good consistency. Now, we test whether such a structure can implement the function of the XOR and OR gate. Generally, the performance of logic gates is evaluated by Extinction Ratio (ER), which is defined as $10\lg(I_1/I_0)$ with $I_1$ ($I_0$) being the output power of logic 1(0). The measurement results for ERs are plotted in Fig. 3b. A detailed description of the measurement method is provided in Appendix B. The corresponding simulation results are shown in Fig. 3c. For simplification, we introduce some symbols to describe logic outputs at XOR and OR ports with different input states. For example, $I_{XOR\_01}$ represents the output intensity of the XOR gate when the input port A is turn-off and the input port B is turn-on. $I_{OR\_11}$ is the output intensity of the OR gate when both input ports A and B are turn-on. Other expressions of logic outputs are based on the same representation method.

Red and blue lines in Figs. 3b and 3c plot the result of $10\lg(I_{OR\_11}/I_{XOR\_11})$ and $10\lg(I_{XOR\_10}/I_{XOR\_11})$, respectively. Due to the existence of mirror symmetry and time-reversal symmetry, the value of $10\lg(I_{XOR\_01}/I_{XOR\_11})$ is the same to that of $10\lg(I_{XOR\_10}/I_{XOR\_11})$. It is seen that the maximum of the ERs is up to 31.78dB, which is much larger than the previous work[31]. We also measure the total transmission, and find that the optical loss is about -0.96dB at λ=1550nm (see S5 of Supporting Materials for details). Moreover, assisted by the multi-wavelength optimization, the operation bandwidth is about 80nm (ER>22dB) in the communication region (1520-1600nm). The experiment results are in good agreement with simulations, indicating a nice performance of XOR and OR gates. It is noted that there is a little difference around 1600nm, which is due to the measurement error resulting from the low coupling efficiency of the 1D grating.

4. The cascaded NAND XNOR NOR AND gates and half adder

To realize the complete set of logic gates, we cascade three logic units designed above (XOR and OR gates), where a 300nm-length waveguide is used to connect each other. Such a cascaded multi-functional logic device is fabricated with its footprint being 1.3×4.5 μm$^2$, as shown in the SEM image of Fig. 4a. There are four input waveguides labeled by BL$_1$, BL$_2$, Input A, and Input B. Input A and Input B are employed to work as the logic input signals. Waveguides (labeled by BL$_1$ and BL$_2$) need to input the bias lights with specific amplitudes when the logic devices are working on. On the right side of the device, three waveguides are set as channels of the logic output. The logic signals of all seven logic gates and a half adder are output from these three channels.

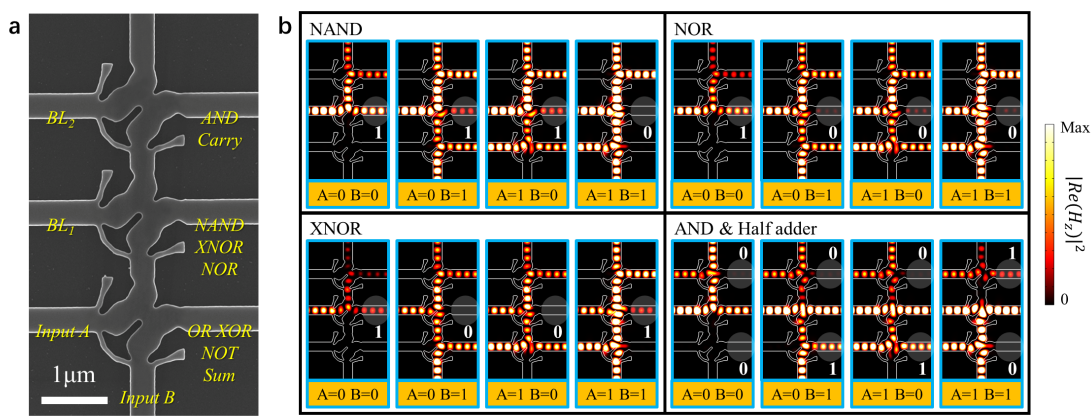

Fig. 4. The cascaded all-optical logic gates and half adder. a, The SEM image of the cascaded device. The corresponding inputs and logic outputs are labeled at the waveguides. b, The simulated steady-state intensity distribution of the logic devices, including NAND, NOR, XNOR, AND gates, and a half adder at λ=1550nm. The logic output of each gate is highlighted in the translucent white circle. Four logic cases, inputs 00, 01, 10, and 11, are listed from left to right in every box.

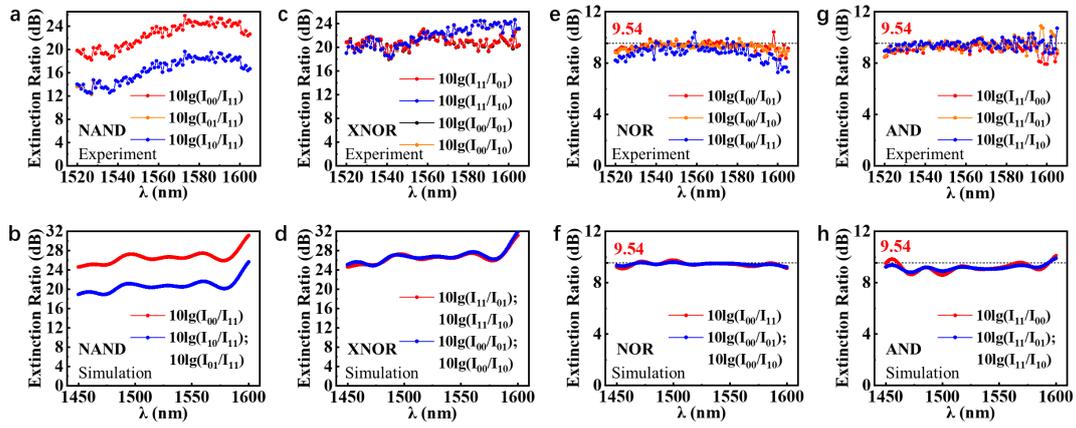

**Fig. 5. The ERs of experiment and simulation for cascaded gates.** The ERs of (**a**) and (**b**) for the NAND gate, (**c**) and (**d**) for the XNOR gate, (**e**) and (**f**) for the NOR gate, (**g**), and (**h**) for the AND gate. (**a**), (**c**), (**e**) and (**g**) are the experiment results; (**b**), (**d**), (**f**) and (**h**) are the simulation results.

Now, we consider three cascaded logic gates, NAND, XNOR, and NOR, whose logic outputs are in the middle output channel marked in Fig. 4a. For the NAND gate, the simulation results of field distributions are shown in the top-left box of Fig. 4b. Four different logic input states are illustrated as (A=0, B=0), (A=0, B=1), (A=1, B=0), and (A=1, B=1), which correspond to four incident cases. The phase difference between Inputs A and B is taken as $\Delta\varphi_{AB}=3\pi/2$ to ensure the interference enhanced logic signal could enter into the next logic unit. At the same time, the bias light at $BL_1$ is always turn-on with the filed amplitude being $\sqrt{2}$ times of that for Inputs A(B). When two input states are turned on (A=1, B=1), the logic 0 could appear at the output channel assisted by the accurate interference cancellation. As for other three cases, the output fields are always nonzero, meaning the NAND gate is realized. Such a phenomenon has also been observed in experiments. In Fig. 5a, we plot experimental results for the ER between the output signal 1 and 0. The maximum is up to 25.82dB, and the operation bandwidth is about 50nm (ER>16dB) in the communication band, which is basically consistent with the simulation results as shown in Fig. 5b.

For the XNOR gate, the excitation conditions are the same to that of the NAND gate, except for the amplitude at $BL_1$ is changed to $\sqrt{2}/2$ times of that for Input A(B). In this case, the ideal destructive interference (the output signal being 0) only appears when one of Inputs A and B is turn-on (A=0, B=1; or A=1, B=0). For other input states (A=0, B=0; and A=1, B=1), the output amplitudes are nonzero. The simulated field distributions are shown in the bottom-left box of Fig. 4b. The experimental results for ERs are plotted in Fig. 5c. It is seen that the maximum for the ER is 24.67dB, and the operation bandwidth is about 80nm (ER>18dB) in the communication band (1520-1600nm), which also corresponds to the simulated results in Fig. 5d. These results show the high efficiency of the XNOR gate.

Furthermore, when the amplitude of the bias light at $BL_1$ is set as $0.75\sqrt{2}$ times of that for Input A(B), the NOR gate can be realized. In particular, when the Inputs A and B are both turn-off (A=0, B=0), only the signal at $BL_1$ contributes to the output power. As for other three cases (A=0, B=1; A=1, B=0; A=1, B=1), the destructive interferences happen at the output channel, and the output powers are significantly reduced. These can be clearly seen in the top-right box of Fig. 4b. The ratio between output powers of the input state of (A=0, B=0) and (A=0, B=1; A=1, B=0; A=1, B=1) is 9:1. Hence, they could be regarded as the logic 1 and 0 with the theoretical ER being 9.54dB. The last one, the AND gate, is realized by adding a NOT gate after the NAND gate. The amplitudes of the bias lights at $BL_1$ and $BL_2$ are set as $\sqrt{2}$ and 0.75 times of that for Input A(B), respectively. Similar to the NOR gate, the theoretical ratio of output powers for the AND gate is also 9:1. The simulated field distributions are shown in the bottom-right box of Fig. 4b. The experiment results of the ER for NOR and AND gates are shown in Figs. 5e and 5g. And, the corresponding simulated results are plotted in Figs. 5f and 5h. A good agreement between experimental and theoretical results is presented.

We would like to emphasize that in order to obtain nice performances of logic gates, the wavelength-dependent phase differences between the bias lights and the input signals need to be suitably designed to ensure the accurate interference cancellation at the output port. In addition, the amplitude of bias lights should multiply a coefficient to compensate transmission losses. The related discussions are given in S6 of Supporting Materials.

In the above integrated logic chip, the XOR, OR and NOT gates can also be realized with only considering Inputs A and B. We note that the phase difference of Input A and B is set as $\Delta\varphi_{AB}=\pi/2$ for the OR gate and $\Delta\varphi_{AB}=3\pi/2$ for XOR and NOT gates. In such a case, XOR and AND gates can be realized simultaneously, i.e. the half adder is also made. The XOR port corresponds to the Sum and the AND port to the Carry. The truth tables of all logic gates and the half adder are listed in Table 1.

**Table 1.** the truth table of the logic gates and the half adder

| Inputs | | All major logic gates | | | | | | | Half adder | |
|---|---|---|---|---|---|---|---|---|---|---|
| Input A | Input B | OR (A+B) | XOR (A⊕B) | NOT ($\bar{A}$) | XNOR ($\overline{A \oplus B}$) | NAND ($\overline{AB}$) | NOR ($\overline{A+B}$) | AND (AB) | CARRY (AND) | SUM (XOR) |
| | | | | | $BL_1=\sqrt{2}/2$ | $BL_1=\sqrt{2}$ | $BL_1=0.75\sqrt{2}$ | $BL_1=\sqrt{2}$ $BL_2=0.75$ | $BL_1=\sqrt{2}$ $BL_2=0.75$ | |
| 0 | 0 | 0 | 0 | | 1 | 1 | 1 | 0 | 0 | 0 |
| 0 | 1 | 1 | 1 | 1 | 0 | 1 | 0 | 0 | 0 | 1 |
| 1 | 0 | 1 | 1 | | 0 | 1 | 0 | 0 | 0 | 1 |
| 1 | 1 | 1 | 0 | 0 | 1 | 0 | 0 | 1 | 1 | 0 |

## 5. Discussion

The full implementation of our scheme in silicon photonics is significant in future optical computing. Our scheme shows a compact multifunctional logic device. Every output port can be connected to the next logic gate by proper layout design. We believe this is a breakthrough in the optical CPU and computer. On the other hand, the proposed logic device can also be used in the optical communication. In particular, the energy consumption per bit is an important index in the optical communication. The power consumption of the logic device is simply given by the total optical input power including BLs. To simplify the analysis, we suppose the bit rate is 20 Gbps[31]. In the experiment, we measure the total optical input power being -19dBm for the XOR/OR gate. Therefore, the energy consumption per bit for the bit rate (20 Gbps) is ~0.6 fJ/bit.

Additionally, the optical phase lock loops (OPLLs) are necessary when our designed logic gates are used. If the OPLLs are applied, it can bring some additional energy consumptions. The above energy consumption per bit may be doubled, because the OPLLs need to expend half input energy to control the phase of the input signal. So, the energy consumption per bit of XOR/OR gates is ~1.2 fJ/bit. For the cascaded gates, the energy consumptions per bit are 1.4fJ/bit (XNOR), 1.8fJ/bit (NOR), 2.4fJ/bit (NAND), and 3.0fJ/bit (AND) when the input powers are taken as -18dBm (XNOR), -17dBm (NOR), -16dBm (NAND), and -15dBm (AND). It can be seen that the energy consumption per bit of our logic device possesses a quite low value. This indicates that a wide application of our designed devices is expected in the optical communication field.

In conclusion, we have designed the ultra-compact all-optical logic devices on the Si photonic platforms using the TO method. The corresponding chips have been fabricated experimentally. With the linear interference approach, all major logic functions (including AND, OR, NOT, NAND, XNOR, XOR, and NOR) and the half adder have been realized. The footprints for a single logic unit (XOR and OR gates) are only 1.3×1.3 μm$^2$ around the optical communication range. Moreover, the integrated chip including 7 major logic gates and a half adder is only 1.3×4.5 μm$^2$. These optimized super-compact all-optical logic devices have an ultra-low loss and a high-efficient performance, which are expected to potentially be applied in future optical information processing.

**APPENDIX A. The comparison table with some reported BSs.**

In Table 2 of this Appendix, we list the main parameters of beam splitters (BSs), directional couplers (DCs), and multi-mode interference (MMI) couplers in previous works. The loss of traditional DC[44, 45] is mainly caused by the waveguide loss. The large footprint leads to a big loss of the DC. As for the compact BS[37, 41-43], and MMI[46] devices, the fabrication error results in device loss. So, the better performance needs a smaller footprint and more precise fabrication technology.

**Table 2.** Comparison table of the reported BSs

| Ref. | Device | Footprint(μm$^2$) | Loss(dB) | Realized method |
|---|---|---|---|---|
| [37] | 1×3 BS | 3.8×2.5μm$^2$ | -0.642dB | Inverse design |
| [41] | 1×2 BS | 2.6×2.6μm$^2$ | -0.46dB* | Deep neural network |
| [42] | 2×2 BS | 4.8×4.8μm$^2$ | about -3dB | Binary particle swarm |
| [43] | 1×2 BS | 1.4×2.3μm$^2$ | -0.5dB | Parameter optimization |
| [44] | 2×2 DC | 40×40μm$^2$ | -2.8dB | Traditional Si wire waveguide |

| [45] | 2×2 DC | 20mm×100μm | -0.7dB/cm | Femtosecond laser writing waveguide |
|---|---|---|---|---|
| [46] | 1×2 MMI | 3.6×11.5μm² | -0.06 dB | Multi-mode interference |
| This work | 2×2 BS | 1.3×1.3μm² | -0.96dB | Topology optimization |

*Only Simulation.

By comparison, we find that our designed device possesses the smallest footprint. In addition, some papers[37, 41, 43] report the lower loss of the 1×3 and 1×2 BSs. However, for the 2×2 BS, the lowest loss is achieved in this work. It is noted that only the 2×2 BS or the DC can be used to realize XOR and OR gates. So, it is significant to design and realize the 2×2 BS with the ultra-small footprint and ultra-low loss.

**APPENDIX B**

**1. Sample fabrication**

The samples were fabricated using electron beam lithography, followed by dry etching. The substrate was a silicon-on-insulator wafer with a 220 nm-thick top Si layer. ZEP-520A e-beam resist was first spin-coated on the substrate for exposure, and resist patterns were formed after e-beam lithography and development. Then these resist patterns were transferred to the top Si layer using inductively coupled plasma etching in $SF_6$ and $CHF_3$ gases atmosphere, with ZEP520A used as an etching mask. The etching depth for logic gates and waveguides is 220nm.

**2. Measurement method**

The continuous wave laser (1520nm-1630nm) was employed to measure output signals of logic gates in the experiment. The incident light was first coupled to the free space from the single mode fiber (SMF). And then we used three polarization beam splitters (PBSs) to split the light into four paths. A series of wave plates were placed in the light path to control the polarization state of lights and the splitting ratio of the PBS. Light beams at those paths were coupled to the SMFs again and entered into the chip by the fiber array. The output signals were collected by another SMF and detected by a high-speed optical power monitor. Phases of beams in four SMFs were unstable because of the temperature variation and eigenvibration of fibers.

We used a dynamic measured method for testing output signals of logic gates. Three fiber phase modulators were used in the measurement, whose modulation periods were set as different values. The detailed experimental set-up and the measurement method are described in S7 of Supporting Materials.

On the other hand, we also propose a scheme to offset the phase unstability by using the OPLLs. In this way, a series of photoelectric devices are used to form a feedback loop for controlling the input phase. The scheme and detailed discussions can be found in S8 of Supporting Materials.

**Supporting Information**

The Supporting Information is available free of charge at https://pubs.acs.org/doi/xxx

The effective permittivity of the Si layer; artificial modification method and its effect on device properties; adjustments of empirical constants; the total transmittance with the length of the side of the design square; the insertion loss of the XOR and OR gates; the phase and amplitude of bias lights; the experimental set-up and the measurement method for the logic device; the scheme of the optical phase lock loops (OPLLs) (PDF)

**Author Contributions**

† L.H. and F. Z. contributed equally.

**Notes**

The authors declare no competing financial interest.

**Funding Sources.**

This work was supported by the National key R; D Program of China under Grant No. 2017YFA0303800, National Natural Science Foundation of China (12004038 and 11904022).

**ACKNOWLEDGMENTS**

L. H. thanks Dr. Rasmus E. Christiansen for helpful discussions about the TO method. The authors thank Prof. Yi Dong, Prof. Wei Zhang, Dr. Yu-Hui Chen, Dr. Pai Zhou for the help in the experiment.